\documentclass[aps,twocolumn,showpacs]{revtex4}
\usepackage{epsfig}
\usepackage{amsbsy}
\usepackage{amsmath}
\usepackage{amsfonts}
\usepackage{graphicx}
\newcommand{\be}{\begin{equation}}
\newcommand{\ee}{\end{equation}}
\newcommand{\rf}[1] {(\ref{#1})}
\def\ket#1{|#1\rangle}
\def\bra#1{\langle #1|}
\newcommand{\Hc}{{\cal H}}

\begin{document}

\title{Entanglement capability of self-inverse Hamiltonian evolution}
\author{Xiaoguang Wang and Barry C. Sanders}
\affiliation{Department of Physics and Australian Centre of Excellence for Quantum Computer Technology, \\
Macquarie University, Sydney, New South Wales 2109, Australia.}

\date{\today}

\begin{abstract}
We determine the entanglement capability of self-inverse Hamiltonian evolution, 
which reduces to the known result for Ising Hamiltonian, and identify optimal 
input states for yielding the maximal entanglement rate. 
We introduce the concept of the operator entanglement rate, and find that the maximal operator entanglement rate gives a lower bound on the entanglement capability of a general Hamiltonian.
\end{abstract}
\maketitle

Two-qubit unitary gates occupy a central role in quantum information science~\cite{Nie00}, 
particularly because of the capability of these gates to generate and enhance entanglement 
of the state of the system. In general, entanglement capability can be enhanced by introducing 
ancillary states~\cite{Dur01,Kra01}, but for the important and interesting case of the Ising 
Hamiltonian $H_{\text{ISING}}=\sigma_z\otimes\sigma_z$ it is ancilla-independent~\cite{Chi02,Childs} (where 
$\sigma_z$=diag$(1,-1)$ is a Pauli matrix). The independence of the entanglement capability 
on ancillas is a consequence of the self-inverse property $H_{\text{ISING}}=H^{-1}_{\text{ISING}}$. 

We generalize this result to all Hamiltonian evolution of the type 
\be
H=X_A\otimes X_B\label{h}
\ee 
such that $X_i=X_i^{-1}\in {\cal H}_i$ for $i\in \{A,B\}$ and $H=H^{-1}$, 
with the Ising Hamiltonian being a special case. Here, we assume that 
$X_i$ is not an identity operator $\openone_i$. Due the self-inverse property of the Hamiltonian, we have the evolution operator ($\hbar=1$)
\be
U(t)=e^{-iHt}=\cos t ~\openone_A\otimes \openone_B-i\sin t ~X_A\otimes X_B.\label{ut}
\ee
We employ operator entanglement~\cite{Zan01,Nie02} and operator entanglement 
rate to characterize the entanglement capability of the self-inverse Hamiltonian evolution: 
these approaches yield elegant analyses that are simple to apply for determining entanglement 
capabilities. We find that the maximal operator entanglement rate 
always gives a lower bound on the entanglement capability of a general Hamiltonian.  

The entanglement capability of a Hamiltonian~\cite{Dur01,Chi02,Kra02} is defined 
relative to a specified entanglement measure. We use von Neumann entropy as our
entanglement measure of a pure state $|\Psi\rangle\in {\cal H}_{AB}$:  
\begin{equation}
E(\ket{\Psi})=-\text{Tr}_A(\rho_A\log\rho_A),\label{entangle}
\end{equation}
where $\rho_A$=$\text{Tr}_B(\rho_{AB})$, $\rho_{AB}=|\Psi\rangle\langle\Psi|$, 
and $\log$ is always base 2. The entanglement capability of Hamiltonian $H$ 
is defined as the maximum entanglement rate when a pure state is acted on by the 
associated evolution operator $ U(t)=\exp(-iHt)$.
Mathematically, the ancilla-unassisted and ancilla-assisted entanglement 
capabilities are defined as~\cite{Dur01,Chi02}
\begin{align}
E_H:&=
\text{max}_{|\Psi\rangle\in{\cal H}_{AB}}\Gamma(t)|_{t\rightarrow 0},\\
E_H^{\text{anc}}:&=
\text{sup}_{|\Psi\rangle\in{\cal H}_{A'ABB'}}\Gamma(t)|_{t\rightarrow 0},
\end{align}
respectively, where 
$
\Gamma(t)={\text{d}E[U(t)|\Psi(0)\rangle]}/{\text{d}t}
$
is the state entanglement rate, and ancilla systems $A'$ and $B'$ are not acted on by Hamiltonian $H$. 
For the ancilla-assisted case the entanglement refers to the bipartite entanglement 
between systems $A'A$ and $BB'$.

Let us first consider the unassisted case, and the initial pure state of 
the system can always be Schmidt-decomposed as~\cite{Peres}
\begin{equation}
|\Psi(0)\rangle=\sum_n \sqrt{\lambda_n}|\psi_n\rangle\otimes|\phi_n\rangle,
\label{psi0}
\end{equation}
where $\{|\psi_n\rangle\}$ and $\{|\phi_n\rangle\}$ are {orthonormal} sets of states, 
and $\lambda_n>0~\forall n$. As $|\Psi(0)\rangle$ is normalized, $\sum_n\lambda_n=1$.
The state at time $t$ is described by the density 
operator $\rho_{AB}(t)=U(t)\rho_{AB}(0)U^\dagger(t)$, which satisfies $ \dot{\rho}_{AB}(t)=-i[H,\rho_{AB}(t)]$. 
Thus, the reduced density operator 
$\rho_A(t)$ satisfies
\begin{equation}
\dot{\rho}_A(t)=-i\text{Tr}_B[H,\rho_{AB}(t)]. \label{dotrho}
\end{equation}
{}From Eq.~(\ref{entangle}), we know that the entanglement rate is ~\cite{Chi02}  
\begin{equation}
\Gamma(t)=-\text{Tr}_A[\dot{\rho}_A(t)\log\rho_A(t)]. \label{dotet}
\end{equation}
Substituting Eq.~(\ref{dotrho}) into Eq.~(\ref{dotet}), we obtain the entanglement rate at $t=0$ as
\begin{equation}
\Gamma(t)|_{t\rightarrow 0}
=i\text{Tr}_A\{\text{Tr}_B[H,\rho_{AB}(0)] \log\rho_A(0)\},\label{eee}
\end{equation}
which is general for arbitrary Hamiltonians and initial states.
By varying initial states we can maximize the entanglement rate.

Equation~\rf{eee} is solved by first obtaining the results
\be
\text{Tr}_B[H,\rho_{AB}(0)]=\sum_{mn}\sqrt{\lambda_m\lambda_n}
(X_B)_{mn}[X_A,|\psi_n\rangle\langle\psi_m|],
\ee
where $(X_B)_{mn}=\langle\phi_m|X_B|\phi_n\rangle$, and
\be
\log\rho_A(0)=\sum_n\log\lambda_n |\psi_n\rangle\langle\psi_n|.
\ee
The entanglement rate at $t=0$ for our Hamiltonian $H$ is then given by,
\be
\Gamma(t)|_{t\rightarrow 0}
=i\sum_{mn}\sqrt{\lambda_m\lambda_n}\log\frac{\lambda_m}{\lambda_n}(X_A)_{mn}(X_B)_{mn},
\label{et0}
\ee
where $ (X_A)_{mn}=\langle\psi_m|X_A|\psi_n\rangle$.

{}From Eq.~(\ref{et0}), we obtain an upper bound for the entanglement rate
\begin{align}
\Gamma(t)|_{t\rightarrow 0}\le& \sum_{mn}\sqrt{\lambda_m\lambda_n}\Big|\log\frac{\lambda_m}{\lambda_n}\Big|~|(X_A)_{mn}|~|(X_B)_{mn}|\nonumber\\
=&\sum_{mn}(\lambda_m+\lambda_n)\sqrt{\frac{\lambda_m}{\lambda_m+\lambda_n}
\frac{\lambda_n}{\lambda_m+\lambda_n}}\Big|\log\frac{\lambda_m}{\lambda_n}\Big|\nonumber\\&\times~|(X_A)_{mn}|~|(X_B)_{mn}|
\nonumber\\
\le&\frac{\beta}{2}\sum_{mn}(\lambda_m+\lambda_n) ~|(X_A)_{mn}|~|(X_B)_{mn}|\nonumber\\
\le&\frac{\beta}{4}\sum_{mn}(\lambda_m+\lambda_n) ~(|(X_A)_{mn}|^2+|(X_B)_{mn}|^2)\nonumber\\
=&\beta,\label{bound}
\end{align}
where 
\begin{align}
\beta=&2\max_{x\in [0,1]}\sqrt{x(1-x)}\log[x/(1-x)] \nonumber\\
=&2\sqrt{x_0(1-x_0)}\log[x_0/(1-x_0)]
\approx 1.9123, \label{betaa}
\end{align}
and $x_0=0.9128$.
The first line simply follows from the triangle inequality, the second and third lines from~Ref.~\cite{Chi02}, and the fourth line from the inequality 
$2|ab|\le |a|^2+|b|^2$ for any complex numbers $a$ and $b$. Finally, the last line results from the self-inverse properties of $X_A$ and $X_B$. To see this, let us examine the sum $\sum_{mn}\lambda_m |(X_A)_{mn}|^2=\sum_{mn}\lambda_m(X_A)_{mn} (X_A)^*_{mn}=\sum_m\lambda_m\sum_n(X_A)_{mn}(X_A)_{nm}=1$, where the second equality results from the Hermitian property of $X_A$, and the last equality from the self-inverse property $X_A=(X_A)^{-1}$.
We have seen that the self-inverse property is essential to obtain the upper bound $\beta$. 
Another feature of Eq.~\rf{bound} is that the result is applicable 
for any pure state with or without ancillas. Therefore, the upper bound 
$\beta$ is {\it ancilla-independent}. Next we show that the upper bound 
can be saturated by optimal input states. 

Since the self-inverse operators $X_i,\, i\in\{A,B\}$ satisfy $X_i^2=1$, the eigenvalues of $X_i$ are $\pm 1$ with the corresponding eigenstates (possibly degenerate) denoted by $|\pm \rangle_{i}$. 
Then, we construct the optimal input states given by~\cite{Referee}
\begin{align}
&|\Psi(0)\rangle=\frac{\sqrt{x_0}}{2}(|+\rangle_{A}+|-\rangle_A)\otimes (|+\rangle_B+|-\rangle_B)\nonumber\\
&+\frac{i\sqrt{1-x_0}}{2}(|+\rangle_A-|-\rangle_A)\otimes (|+\rangle_B-|-\rangle_B).
\end{align}
This state is of Schmidt form with Schmidt number 2. From Eqs.~(\ref{et0}) and (\ref{betaa}), we obtain $\Gamma(t)|_{t\rightarrow 0}=\beta$. Therefore, we can always find the optimal input states for which 
\be
E_H=E_H^{\text{anc}}=\beta.\label{eh}
\ee
Result\,\rf{eh} extends results for the Ising Hamiltonian~\cite{Dur01,Chi02}, and we do not restrict the dimension of Hilbert spaces on which the self-inverse operators $X_A$ and $X_B$ act. The Hilbert space can be finite-dimensional or infinite-dimensional.

Self-inverse Hamiltonians not only 
exhibit elegant entanglement capability, but are also physically meaningful. 
As an example of a physical system with optimal input state and maximal entanglement rate $\beta$, 
we consider a spin-$j$ system with a $(2j+1)$-dimensional space spanned by basis 
states $\{|n\rangle_j\equiv |j;n-j\rangle, n=0,1,\ldots,2j\}$. We define a 
number operator ${{\cal N}}=J_z+j$ of the spin-$j$ system satisfying ${{\cal N}}|n\rangle_j=n|n\rangle_j$. 
Here, commutators for $J_z$ and ladder operators $J_{\pm}$ satisfy the su(2) algebra $[J_z,J_{\pm}]=\pm J_{\pm}, ~[J_+,J_-]=2J_z$.
Then, using the number operators, we construct the self-inverse interaction Hamiltonian
\be
H_1=(-1)^{{{\cal N}}}\otimes (-1)^{{\cal N}},\label{h1}
\ee
where $(-1)^{{\cal N}}$ is a parity operator of the system. 
This kind of Hamiltonian is realizable 
in physical systems~\cite{Geometric}. {}From Eq. (\ref{h1}),  we see that Hamiltonian $H_1$ reduces to the Ising Hamiltonian $H_{\text{ISING}}$ for the case of $j=1/2$. When the dimension $d=2j+1$ is even, i.e., $j$ is a half integer, Hamiltonian $H_1$ is equivalent to the Ising Hamiltonian in the sense that the operator $(-1)^{\cal N}$ can be written as $\openone_{d/2}\otimes \sigma_z$, where 
$\openone_{N}$ denotes the $N\times N$ identity matrix. For integer $j$, Hamiltonian $H_1$ is not equivalent to 
the Ising Hamiltonian.
 
Let us consider the SU(2) entangled coherent state~(ECS)~\cite{ECS} 
$|\eta\rangle_{\text{ECS}}$ as an input state 
\be
|\eta\rangle_{\text{ECS}}=\sqrt{x_0}|\eta\rangle\otimes|\eta\rangle
+i\sqrt{1-x_0}|-\eta\rangle\otimes|-\eta\rangle, \label{ecs}
\ee
where 
$|\eta\rangle=\exp(\eta J_+-\eta^* J_-)|0\rangle_j$ is the SU(2) or spin coherent state~(SCS)~\cite{SCS}, 
$\eta$ is complex with unit modulus. 
{}From Eq.~(\ref{et0}) and the identity $(-1)^{{\cal N}}|\eta\rangle=|-\eta\rangle$, it is straightforward to show that the SU(2) ECS is the
optimal input state which yields the maximal entanglement rate $\beta$.

The above analysis is restricted to finite equal-dimensional composite systems, but this limitation is convenient, not essential. 
Consider the harmonic oscillator with infinite-dimensional Hilbert space.
The operator $a^\dagger a$ is the number operator, where $a^\dagger$ ($a$) is  the creation (annihilation) operator.
We can have the Hamiltonians: $H_2=(-1)^{{{\cal N}}}\otimes (-1)^{a^\dagger a}$ and $H_3=(-1)^{a^\dagger a}\otimes (-1)^{a^\dagger a}$. 
By noticing that the SCS can be realized in the Fock space as the binomial state~\cite{BS}, the optimal input states for Hamiltonians $H_2$ and $H_3$ are directly obtained by appropriately replacing SCS with the binomial state in the SU(2) ECS~\rf{ecs}.

Although in general one cannot expect an entangled state to be generated by Hamiltonian evolution from 
a product state, in this spin system the optimal input state $|\eta\rangle_{\text{ECS}}$ 
can be generated by the Hamiltonian $-H_1$ from the product state $|\eta\rangle\otimes|\eta\rangle$. 
Mathematically, $|\eta\rangle_{\text{ECS}}=\exp(iH_1t)|\eta\rangle\otimes|\eta\rangle$ 
with $x_0=\cos^2t$.
This fact and the ancilla-independence of the self-inverse Hamiltonian 
entanglement capability suggest that the entanglement capability depends more on the Hamiltonian and the associated unitary operator. Next, we demonstrate this 
character by introducing the entanglement rate for 
operator $U(t)$, and give the result that the entanglement 
capability of Hamiltonian $H_1$ is equal to the maximal operator entanglement rate for the associated unitary operator. 

We review operator entanglement introduced in Ref.~\cite{Zan01}.  
Notice that the linear operators over ${\cal H}_d$ (finite $d$) also form a $d^2$-dimensional Hilbert space denoted by ${\cal H}_{d^2}^{\text{HS}}$, and the corresponding scalar product between two operators $X$ and $Y$ is given by
the Hilbert-Schmidt product $\langle X,Y\rangle :=\text{Tr}(X^{\dagger }Y)$, and $||X||_{\text{HS}}=\langle X,X\rangle.$ 
Then, the operator acting on ${\cal H}_{d_1}\otimes {\cal H}_{d_2}\,$can be considered as a state on the composite Hilbert space ${\cal H}_{d_1^2}^{\text{HS}}\otimes {\cal H}_{d_2^2}^{\text{HS}}$, and the operator entanglement can be defined as the entanglement of that state.

Any unitary operator $V$ acting on ${\cal H}_{d_1}\otimes {\cal H}_{d_2}\,$ may be Schmidt-decomposed as~\cite{Nie02} $V=\sum_ns_nA_n\otimes B_n$, 
where $s_n> 0~\forall n $ and $A_n$ and $B_n$ are {orthonormal} operator bases for
systems $1$ and $2.$ From the Schmidt form the entanglement of a unitary
operator $V$ is determined to be
$
{\cal E}(V)=-\sum_l{s_l^2}/{(d_1d_2)}
\log\left[{s_l^2}/{(d_1d_2)}\right],
$
where the factor $1/(d_1d_2)$ arises from normalization of the unitary operator. 
We can think of the operator entanglement as a strength measure of the operator~\cite{Nie02}.

If we consider the unitary operator $U(t)$, the operator entanglement ${\cal E}[U(t)]$ becomes a time-dependent function. Analogous to the definition of the state entanglement rate~\cite{Dur01}, it is natural to define the operator entanglement rate at a certain time of interaction and the maximal entanglement rate
\begin{equation}
R(t):={\text{d} {\cal E}[U(t)]}/{\text{d}t}, \quad R_{\max}:=\max_t R(t),
\end{equation}
respectively, where the maximization is over all time $t$.  At certain times the 
entanglement rate becomes maximal. 

The operator entanglement of an 
arbitrary unitary operator $V$ is equal to 
the entanglement of the 
state $V(|\Phi \rangle _{A^{\prime }A}\otimes |\Phi \rangle _{BB'}) $~\cite{Nie02}, 
\begin{equation}
{\cal E}(V)=E[V(|\Phi \rangle _{A'A}\otimes |\Phi \rangle _{BB'})], \label{opstate}
\end{equation}
where
\begin{align}
|\Phi \rangle _{A^{\prime }A} &=\sum_{n=0}^{d_1-1}|n\rangle _{A^{\prime
}}\otimes |n\rangle_A 
\, \in {\cal H}_{d_1}\otimes {\cal H}_{d_1},\\
|\Phi \rangle _{BB'} &=\sum_{n=0}^{d_2-1}|n\rangle _B\otimes
|n\rangle _{B^{\prime }}
\, \in {\cal H}_{d_2}\otimes {\cal H}_{d_2}
\end{align}
are maximally entangled states. 
Equation (\ref{opstate}) shows a direct relation between operator entanglement and state entanglement. 
Note that here dimension $d_1$ can differ from dimension $d_2$.
{}From Eq.~\rf{opstate}, we immediately have a relation between the operator entanglement rate and the state entanglement rate
\begin{equation}
R(t)=\Gamma(t)=\dot{E}[U(t) (|\Phi \rangle_{A'A} \otimes |\Phi \rangle_{BB'})].\label{rrr}
\end{equation}
From Eq.~\rf{rrr} we find that {the maximal operator entanglement rate gives a 
lower bound for the entanglement capability of a general Hamiltonian}, which is the infinitesimal version of the one that {the operator entanglement gives a lower bound of the entanglement capability of a unitary operator}~\cite{Nie02}.

Having defined the operator entanglement rate let us study the maximal operator entanglement rate $R_{\max}$ of the evolution operator associated with Hamiltonian $H_1$ with dimension $d=2j+1$ an even number~\cite{Even}. The associated unitary operator is in the Schmidt form
\be
U_1(t)=\cos t~\openone_d\otimes \openone_d -i\sin t~(-1)^{\cal N}\otimes (-1)^{\cal N}.\label{u1t}
\ee
{}From the above equation, the operator entanglement and operator entanglement rate are given by
\begin{align}
{\cal E}[U_1(t)]=&-\cos^2t\log\cos^2t-\sin^2t\log\sin^2t,\\
R(t)=&R[U_1(t)]=\sin(2t)\log(\cot^2 t),
\end{align}
respectively.
The above equation shows that the operator entanglement rate is a periodic function of time $t$; 
hence we can maximize over one period. It is straightforward to find that 
at time $t=0.2932$, the entanglement rate reaches its maximum value, i.e., 
\be
R_{\max}=\beta.\label{r}
\ee
Comparing Eqs.~(\ref{eh}) and (\ref{r}), we conclude that the entanglement capability 
of Hamiltonian $H_1$ is equal to the maximum operator entanglement rate of the associated 
unitary operator $\exp(-iH_1t)$. Therefore, the entanglement capability of the self-inverse
Hamiltonian is inherent in the evolution operator in the sense that it can be solely determined by the evolution operator, irrespective of states.

More generally, there is interest in the 
entanglement capability of quantum operators, not just Hamiltonians~\cite{Mak00,Zan00,Kra01,Lei02,Ben02,Ber02,Wol02,WanSan02}.
Here, we analyze the entanglement capability of the unitary operator 
generated by the self-inverse Hamiltonian $H$ \rf{ut}.
We quantify the entanglement capability of a unitary operator $U$ 
by the maximum entanglement which a unitary operator can create given an initial product state~\cite{Kra01}:
\be
E_U=\text{sup}_{\ket{\gamma},\ket{\delta}} E(U\ket{\gamma}\otimes \ket{\delta}),
\ee
where $\ket{\gamma}\in {\cal H}_{A'A}$ and $\ket{\delta}\in {\cal H}_{BB'}$, namely, we include ancillas.
For unitary operator $U(t)$ we immediately obtain
\begin{align}
E_{U(t)}=&\text{sup}_{\ket{\gamma},\ket{\delta}} E[U(t) \ket{\gamma}\otimes \ket{\delta}]
=\text{sup}_{\ket{\gamma},\ket{\delta}} E[|\Psi(t)\rangle],
\\
\ket{\Psi(t)}=&\cos t ~\ket{\gamma}\otimes \ket{\delta}-i\sin t ~\ket{\bar{\gamma}}\ket{\bar{\delta}},\label{ssstate}
\end{align}
with normalized states $\ket{\bar{\gamma}}=X_A\ket{\gamma}$ and $\ket{\bar{\delta}}=X_B\ket{\delta}$.

The state (\ref{ssstate}) belongs to a class of bipartite entangled states discussed in Ref.~\cite{Non}. 
To quantify the entanglement of this state, we can use the standard entanglement measure, the entropy of entanglement. However, for convenience, we use the concept of the concurrence $C$~\cite{Conc} to quantify the entanglement since the state can be viewed as a state of two pseudo-qubits~\cite{Wangjpa}. 
Using the results of Ref.~\cite{Wangjpa}, 
the concurrence of state $\ket{\Psi(t)}$ is obtained as
\begin{equation}
C[\ket{\Psi(t)}]=|\sin(2t)|\sqrt{
(1-|\bra{\gamma}\bar{\gamma}\rangle|^2)
(1-|\bra{\delta}\bar{\delta}\rangle|^2)}\label{cc}
\end{equation}
{}From the above equation, it is easy to find that 
$
E_{U(t)}= |\sin(2t)|,
$
which is the maximal entanglement which the operator $U(t)$ can generate 
from an arbitrary product state. As the states $\ket{\gamma}\in {\cal H}_{A'A}$ and $\ket{\delta}\in {\cal H}_{BB'}$, 
the entanglement capability $E_{U(t)}$ is ancilla-independent. 

The unitary operaror $U_1(t)$ (\ref{u1t}) is in fact a special case of $U(t)$ (\ref{ut}).
Assuming that $U_1(t)$ act on $\Hc_d\otimes \Hc_d$ with even $d$, we find that
the optimal input state for generating maximal entanglement is $\ket{\eta}\otimes\ket{\eta}$, and 
from Eq.~\rf{u1t} we find that 
\be
{\cal E}[U_1(t)]=C[U_1(t)]=E_{U_1(t)}=|\sin(2t)|.
\ee
Therefore, the operator entanglement of $U_1$ is equal to the entanglement capability of the unitary operator.

In conclusion, we have determined the entanglement capability of self-inverse Hamiltonian evolution. 
The self-inverse Hamiltonians studied here are physically relevant
in their own right, and reduce to the important Ising Hamiltonian that is central to quantum information science. 
These Hamiltonians go beyond the two-qubit cases; namely, they can act on $d_1\times d_2$ composite system, where the dimension of the subsystem can be either finite or infinite.
We introduced the concept of operator entanglement rate which is well-defined for composite finite systems, and for certain self-inverse Hamiltonians the maximal operator entanglement rate is equal to the entanglement capability.  
For a general Hamiltonian the maximal operator entanglement rate always gives a lower bound on the Hamiltonian 
entanglement capability.

{\bf Acknowledgments}: The authors acknowledge helpful discussions 
with Paolo Zanardi, Dominic Berry, Stephen Bartlett, and Gerard Milburn. This project has been supported by an Australian Research Council Large Grant.


\begin{thebibliography}{}
\bibitem{Even}The even number $N$ guarantees that the Hilbert-Schmidt inner product of the identity operator and $(-1)^{\cal N}$ is zero. Thus, the operator 
$\exp(-iH_1t)$ is of Schmidt form.
\end{thebibliography}
\end{document}